\documentclass[aps,prb,reprint]{revtex4-1}
\usepackage{graphics}
\usepackage{graphicx}
\usepackage{amssymb}
\usepackage{color}
\begin{document}
\newcommand{\RR}{\mathrm{\mathbf{R}}}
\newcommand{\rr}{\mathrm{\mathbf{r}}}
\newcommand{\defin}{\stackrel{def}{=}}
\renewcommand{\thetable}{\arabic{table}}
\renewcommand{\thefigure}{\arabic{figure}}

\title{Donors in Ge as Qubits -- Establishing Physical Attributes}

\author{A. Baena}

\affiliation{Instituto de F\'{\i}sica, Universidade Federal do Rio de Janeiro, C.P. 68528, 21941-972 Rio de Janeiro, Brazil}

\author{A. L. Saraiva}

\affiliation{Instituto de F\'{\i}sica, Universidade Federal do Rio de Janeiro, C.P. 68528, 21941-972 Rio de Janeiro, Brazil}

\author{Marcos G. Menezes}

\affiliation{Instituto de F\'{\i}sica, Universidade Federal do Rio de Janeiro, C.P. 68528, 21941-972 Rio de Janeiro, Brazil}

\author{Belita Koiller}

\affiliation{Instituto de F\'{\i}sica, Universidade Federal do Rio de Janeiro, C.P. 68528, 21941-972 Rio de Janeiro, Brazil}

%\pacs{03.67.Lx}{Quantum computation architectures and implementations}
%\pacs{73.22.Dj}{Single particle states}
%\pacs{07.79.Cz}{Scanning tunneling microscopes}

\date{\today}

\begin{abstract}
Quantum electronic devices at the single impurity level demand an understanding of the physical attributes of dopants at an unprecedented accuracy. Germanium-based technologies have been developed recently, creating a necessity to adapt the latest theoretical tools to the unique electronic structure of this material. We investigate basic properties of donors in Ge which are not known experimentally, but are indispensable for qubit implementations. Our approach provides a description of the wavefunction at multiscale, associating microscopic information from Density Functional Theory and envelope functions from state of the art multivalley effective mass calculations, including a central cell correction designed to reproduce the energetics of all group V donor species (P, As, Sb and Bi). With this formalism, we predict the binding energies of negatively ionized donors (D$^-$ state). Furthermore, we investigate the signatures of buried donors to be expected from Scanning Tunneling Microscopy (STM). The naive assumption that attributes of donor electrons in other semiconductors may be extrapolated to Ge is shown to fail, similar to earlier attempts to recreate in Si qubits designed for GaAs. Our results suggest that the mature techniques available for qubit realizations may be adapted to germanium to some extent, but the peculiarities of the Ge band structure will demand new ideas for fabrication and control.
\end{abstract}

\maketitle

\section{Introduction} \label{sec:intro} Single donor electronics, fully governed by quantum mechanics, is now accessible due to progress in sample fabrication and characterization~\cite{Simmons12,Rogge13,Zwanenburg13,koenraad-natmat-2011}. Advances in the study of isolated donors in Si naturally motivate investigations in Ge along similar paths~\cite{scappucci2015bottom}. Atomically precise fabrication of devices based on STM litography have been demonstrated \cite{scapucci2011}, paving the road for future applications such as quantum computation. Indeed, germanium shares many of the interesting properties of Si, such as the low abundance of isotopes with nuclear spin in the natural material. This leads to microseconds long coherence of electron spins~\cite{sigillito2015}, which has been extended up to milliseconds in isotopically purified samples. More recent results also point towards feasible Stark tuning of the electron spin resonance, which may serve as a scheme for individual dopant control~\cite{sigillito2016}. But unlike silicon-based technologies, current knowledge about the physics of electrons in Ge is limited. Experimental determination of negatively-charged ionic states-- known to play an important role in nanoscopic devices -- is lacking. The theoretical description of the electronic wavefunction at the atomic level is also incomplete.

We present here a comprehensive investigation for isolated donor-based qubits in Ge with one and two electrons. Our model incorporates parameters which consistently reproduce and predict specific fingerprints for substitutional P, As, Sb and Bi dopants. Calculated ground state energies for neutral ($D^0$)and negatively ionized ($D^-$) donors, binding and charging energies are in good agreement with experiments; and when experimental values are not available, suitable estimates are given. Combining multivalley effective-mass Kohn-Luttinger (KL) envelopes~\cite{kohn55} and ab-initio calculated Bloch functions at the $L$ points, we simulate sub-surface donor images to be expected in STM experiments on a [001] surface, providing additional probes on the presence and location of sub-surface donors in Ge. The multiscale nature of our approach is made necessary by the variety of length scales that appear in this problem, ranging from the sub-nanometer central cell potential to the loosely confined wavefunction of the ionic state, which spreads over tens of lattice parameters.

Recent donor image studies in Si~\cite{STM-si} reveal the unsuspected adequacy of the KL ground state wave function to describe donors' signatures probed at a surface a few monolayers above it. The prominent role played by the host material atomic arrangement and conduction band structure is also discussed in Ref. \onlinecite{STM-si}. Although Ge is also a group IV element and shares the same crystal structure of Si (the diamond structure), the conduction band minima of Ge lie at 4 nonequivalent L points. This leads to dopant signatures in Ge that are very distinct from those of Si. Density Functional Theory calculations provide plane wave expansion coefficients for the Bloch functions of a set of four inequivalent $L$-points.

%The article is organized as follows: in the next section, we present the formalism used in this work along with a discussion of the neutral donor energetics. In section \ref{sec:results}, we present our results for the energetics of the charged donor states and for the subsurface signatures of neutral donors in Ge. In section \ref{sec:discussions} we provide a discussion of the results and in section \ref{sec:conclusions} we present our conclusions.

\section{Formalism} \label{sec:formalism}

The original effective mass equation for shallow donors in Ge incorporates the conduction band mass anisotropy ($m^{\rm Ge}_L = 1.58, m^{\rm Ge}_T=0.082$ in units of the bare electron mass $m_0$) and a screened Coulomb potential ($\epsilon_{\rm Ge}=15.36$), but not the band edge degeneracy~\cite{BSGe}. For Ge, within the Kohn-Luttinger single valley approach (KLSV), the calculated energy of an electron in the  $D^0$ ground state is $E_{\rm KLSV}=-9.05$ meV, regardless of the donor species. For energetics, we may further simplify the single valley formalism by assuming the band is isotropic, characterized by a single mass consistently chosen so that the ground state energy equals the KLSV value, i.e. $E^* = m^*_{\rm Ge}/(\epsilon_{\rm Ge})^2 \times (-13.6 \rm{ eV}) = E_{\rm KLSV}$ which gives   $m^*_{\rm Ge} = 0.156$ in units of $m_0$. This calculation also leads to the hydrogenic approximation for the wavefunction radius $a^*=5.22$ nm. On the other hand, for wavefunction calculations below we take the mass anisotropy into account.

Experimentally, the ground state binding energies of group V donors in Ge are larger than $|E_{\rm KLSV}|$, ranging from $10$ to $14$ meV. This is expected, since inter-valley coupling mediated by the impurity perturbation potential $V_{\rm IMP}$ is not explicitly included in the single valley formalism. Within the isotropic band assumption, we take the impurity potential as $V_{\rm IMP}(r) =-e^2/\varepsilon_{\rm Ge} r+V_{\rm CC}(r)$, in which we include a central-cell contribution $V_{\rm CC}(r)$ which correctly reproduces the behavior of screening for $r\to 0$ and $r\to \infty$, given by
\begin{equation}
V_{\rm CC}(r)= -\bigg ( 1- \frac{\varepsilon_{\rm 0}}{\varepsilon_{\rm Ge}} \bigg  ) e^{-r/r_{cc}} \frac{e^2}{\varepsilon_{\rm 0}r}.
\label{eq:Vcc}
\end{equation}

The donor-specific empirical parameter $r_{\rm cc}$ is a crossover length between the short and long-range behavior of the potential and it completely defines the central-cell-corrected hamiltonian. It is chosen from a variational calculation by assuming hydrogenic envelopes of the form $F_{\rm H}(r,a_{\rm cc})=({ \pi a_{\rm cc}^{3}})^{-1/2}e^{-r/a_{\rm cc}}$, taking the effective Bohr radius $a_{cc}$ as the variational parameter. We take $r_{\rm cc}$ so that  the variational and experimental ground state energies match:  $E^0_{\rm calc}(a_{\rm cc},r_{\rm cc})  = E^0_{\rm exp}$.

In Table \ref{table-D0}, we give the values of  $r_{cc}$ and $a_{cc}$ for P, As, Sb and Bi donors at the neutral state in Ge. Note that $r_{cc}$ is over one order of magnitude smaller than $a_{cc}$ for all donors, thus indicating that the central cell correction is only relevant within a narrow region around the donor site.

The effects of the intervalley coupling on the energetics of neutral donors can be incorporated in our single valley formalism in the following way. Within the KL multivalley effective mass theory \cite{kohn55}, the wavefunction of a neutral donor in Ge can be described in terms of $1$s-like envelopes combined with the Ge Bloch functions at the four band minima:
\begin{equation}
\Psi(r)= \sum_{\mu=1}^4 \alpha_\mu F_{\mu}(\mathbf{r}) e^{ i\mathbf{ k \cdot r}} u_{\mu}(\mathbf{r})= \sum_{\mu=1}^4 \alpha_\mu \phi_{\mu}(\mathbf{r})~,
\label{eq:WF}
\end{equation}

\noindent where the index $\mu$ runs over four nonequivalent L points and $ u_{\mu}(\mathbf{r})$ denotes the periodic part of the corresponding Bloch function. Intervalley coupling in Ge lifts the fourfold degeneracy (energy $E_{KLSV}$) into a non-degenerate ground state $1s (A_{1})$ and three degenerate excited states $1s (T_{2})$. Each of these states is characterized by different coeficients $\{\alpha_\mu\}$ on Eq. (\ref{eq:WF}).

It can be easily verified that the energies of the $A_1$ and $T_2$ states may be written as $E(A_{1})= E_{KLSV} - \delta +12\Delta$ and $E(T_{2})=E_{KLSV} - \delta- 4\Delta$, where $-\delta = \langle \phi_\mu | V_{\rm IMP} (r)|\phi_\mu \rangle $ is the first order correction to the single valley energies and $\Delta = \langle \phi_\mu | V_{\rm IMP} (r)|\phi_{\nu \neq \mu} \rangle $. The two shift parameters $\delta$ and $\Delta$ are obtained by matching the theoretical and experimental values of $E(A_{1})$ and $E(T_{2})$, thus allowing the incorporation of the energy splitting into our description, where only one isotropic envelope $F(r)$ is involved.  Table \ref{table-D0} shows the shift and central cell parameters, and the energies of the $A_1$ and $T_2$ states for neutral As, P, Sb and Bi  donors. The experimental and theoretical results in this table are, therefore, designed to match accurately. From these results we can make new predictions, as we discuss below.

\begin{table}[ht!]
\centering
\begin{tabular}{|c |c| c| c| c| }
\hline
D$^{0}$ &  \ \ \ \  P$^{0}$ \  \  \  \   &   \  \ \ \  As$^{0}$  \  \  \  \  & \ \ \ \   Sb$^{0}$ \ \ \ \   & \  \  \  \  Bi$^{0}$   \ \ \ \  \\
\hline
 $ \delta $ & $1.71$ & $1.95$ & $1.05$ & $1.02$ \\
$\Delta $ & $-0.17$ & $-0.26$ & $-0.02$ & $-0.20$ \\
\hline
a$_{cc}$ & $2.89$ &  $2.33$   & $4.19$  &   $2.97$  \\
  r$_{cc}$ & $0.24$  & $0.26$ & $0.18$ & $0.24$ \\
\hline
A$_{1}^{\rm calc}$& $-12.87$ & $-14.18$ & $-10.44$ & $-12.70$ \\
T$_{2}^{\rm calc}$ & $-10.06$ & $-9.94$ & $-9.95$ & $-9.97$ \\
\hline
A$_{1}^{\rm exp}$& $-12.88$  & $-14.18$ & $-10.45$  & $-12.75$ \\
T$_{2}^{\rm exp}$& $-10.06$ & $-9.94$ & $-9.99$ & $-9.90$\\
 \hline
\end{tabular}
\caption{Parameters for the shift ($\delta$ and $\Delta$) and central cell ($r_{cc}$ and $a_{cc}$), and energies of the $A_1$ and $T_2$ states for neutral donors in Ge. All energies are in meV and lenghts in nm.}
\label{table-D0}
\end{table}

\section{Binding and Charging Energies}

We turn our attention to the negatively charged donor state. Here we follow an analogy with the H$^-$ and He problems in atomic physics~\cite{Bethe-Salpeter}, as described in Ref.~\cite{Koiller2014}. Within the isotropic single valley model, the two-electron Hamiltonian is written as:
\begin{equation}
H_{D^-}=K_{1}+K_{2}+V_{\rm IMP}^{1}+V_{\rm IMP}^{2}+e^2/(\epsilon_{\rm Ge} r_{12}),
\label{eq:Dm}
\end{equation}
\noindent where $K_i$ is the kinetic energy operator for electron $i$, $V_{\rm IMP}^i$ is the central-cell corrected donor potential for electron $i$ and the last term is the Coulomb repulsion between the electrons. For the impurity  potentials, we use the same expressions and values of $r_{cc}$ obtained from the neutral donor state calculations, as given by Table \ref{table-D0}.

The $D^-$ energy is obtained variationally by assuming that the two-electron wavefunction is a symmetric combination of  $1s$ envelopes~\cite{Bethe-Salpeter}
\begin{equation}
\Psi_{D^-}  =   \psi(r_1,a_{\rm i})\psi(r_2,a_{\rm o})   +   \psi(r_1,a_{\rm o})\psi(r_2,a_{\rm i}),
\end{equation}
where $a_{\rm i}$ and $a_{\rm o}$, represent the effective Bohr radii of inner and outer orbitals, taken as variational parameters. The  values of $a_{\rm i}$, $a_{\rm o}$ and of the variational energy $E^-$ are given in Table \ref{table-D-}.

Notice that $a_{\rm o}$ may be as large as 16 nm and is strongly impacted by the central cell potential, which is present at the scale of $r_{cc}\approx 0.2$ nm. To describe physical attributes at such wide range of length scales requires a multiscale description such as the one developed here.

From the energy $E^-$ and the energy of the $D^0$ state $E^0$, we estimate the $D^-$ binding and charging energies as $E_{B}=E^0 - E^-$  and $U = E^{-} - 2 E^0$, respectively. Those energies are also reported in Table \ref{table-D-}, together with experimental data, available only for As$^-$  and Sb$^-$. Since our model for the $D^-$ state does not include any additional empirical parameters, the good agreement with available experiments certifies the consistency of our model, thus indicating that the values for $E_B$ and $U$ for P$^-$ and Bi$^-$ in Table \ref{table-D-} are probably reliable estimates.
\begin{table}[ht!]
\centering
\begin{tabular}{|c |c| c| c| c| }
\hline
 D$^{-}$ & P$^{-}$&As$^{-}$&Sb$^{-}$&Bi$^{-}$ \\
 \hline
 a$_{\rm i}$ & $2.76$ &  $2.14$   & $4.04$  &   $2.81$  \\
  a$_{\rm o}$& $12.89$  & $10.46$ & $16.74$ & $13.16$ \\
\hline
 E$_{\rm Exp}^{-}$ & N$\setminus$A &  $-14.93$   & $-11.07$   & N$\setminus$A   \\
 E$_{\rm Calc}^{-}$ &  $-13.22$  & $-14.71$ & $-10.69$  & $-13.07$ \\
\hline
 E$_{\rm B_ {\rm Exp}}$ &   N$\setminus$A & $0.75$   & $0.62$   & N$\setminus$A   \\
 E$_{\rm B_ {\rm Calc}}$ &  $0.35$  & $0.53$ & $0.25$  & $0.37$ \\
\hline
 U$_{\rm Exp}$ &   N$\setminus$A & $13.43$   & $9.83$   & N$\setminus$A   \\
 U$_{\rm Calc}$ &   $12.52$ & $13.65$   & $10.19$   & $12.33$   \\
 \hline
\end{tabular}
\caption{Parameters  a$_{\rm i}$,  a$_{\rm o}$ and  $E^-$ as obtained variationally for negatively ionized donors (P, As, Sb and Bi)
in Ge. All lengths are in nm and energies in meV. The estimated binding ( $E_{\rm B}= \rm E^{0} - \rm E^{-}$) and charging  ($\rm U=\rm E^{-} - 2\rm E^{0})$ energies are given as well as their experimental values \cite{Opticalabsbook}.  No experimental values are available for P and Bi in Ge.}
\label{table-D-}
\end{table}

\section{Subsurface Donor Signature} Fingerprints of subsurface donors​ in semiconductors ​ are now accessible from STM images \cite{koenraad-natmat-2011}. Recent scans on a (001) surface in Si \cite{STM-exp14} ​identify the presence of buried donors, with a charge distribution aligned along the direction defined by the surface dimerization. The overall appearance of the STM images of donors in Si were satisfactorily described by a model that assumes that the tunnel current is proportional to the local electronic density of states; the STM tip is kept at a constant height from the surface; and the electronic charge distribution near a surface shows negligible changes due to the presence of the surface, keeping the corresponding bulk features. In fact the simulated images adopted in comparison to the STM images are merely two-dimensional cuts of the KL multivalley density through interstitial lattice planes \cite{STM-si}. Similar measurements are being performed in Ge:P with increasing level of refinement~\cite{scappucci2009influence,mattoni2013phosphorus}, leading to the expectation that a complete profile of the impurity charge density will soon be available. For this reason we explore, within the same assumptions, images to be expected from such measurements in Ge. We may also adopt the dimer row direction as a reference for the alignment of charge distributions in Ge.

For the Ge donor $A_1$ ground state, the general expression in Eq. ({\ref{eq:WF}}) takes coefficients $\alpha_\mu=1/2$ for all $\mu$.
The envelope functions $F_{\mu = 1,\ldots 4}({\mathbf r})$ will now include the effects of mass anisotropy. We choose the cartesian axis along the three cubic crystal directions in Ge. For a simple donor, the envelope for the $L_\mu$ minimum along $[\ell_x^\mu,\ell_y^\mu, \ell_z^\mu]$,  with $\ell_{x,~y,~z}^\mu = \pm 1$, is written as
\begin{eqnarray}
F_{\mu}(\mathbf {r})&=&N \exp - \bigg [\frac{(x'+y'+z')^2}{3a_{L}^2} \nonumber \\
 &+& \frac{(x'-y')^2}{2a_{T}^2}+\frac{(-x'-y'+2z')^2}{6a_{T}^2} \bigg ]^{1/2}
\label{eq:envelopes}
\end{eqnarray}
\noindent where $\mathbf {r}=(x,y,z)=(\ell_x^\mu\times x',\ell_y^\mu\times y', \ell_z \times z') $, $a_{L}$ and $a_{T}$ are the longitudinal and transverse effective Bohr radii and $N$ is the normalization constant. By using the experimental data from Ref. \onlinecite{madelung}, we obtain $a_{L}=2.157 \rm nm$;~$a_{T}=6.069\rm nm$ for Ge. Central-cell corrected values are estimated by rescaling those values by the same factor obtained in the spherical approximation above (e.g., for a P donor the scaling factor is $a_{cc}/a^*=2.89/5.22=0.55$).

Following the usual approach in ab-initio calculations, we identify the Bloch wavefunctions with the corresponding Kohn-Sham wavefunctions from Density Functional Theory~\cite{PhysRev64,kohn-sham65}. They were obtained here from an \textit{ab-initio} calculation using the Quantum Espresso code~\cite{Gianozzi09}. The periodic part of the Bloch wavefunction $u_{\mu}({\bf r})$ is expanded in plane waves as $u_{\mu}({\bf r}) = \sum_{\bf G} c_{\mu} ({\bf G}) e^{i\bf G\cdot r}$, where the summation runs over all reciprocal lattice vectors ${\bf G}$ up to an energy cutoff of $80.0$ Ry. \footnote{In the DFT calculation, we use a PZ-LDA exchange-correlation functional for the electron-electron interactions~\cite{perdew81}. The Brillouin Zone is sampled with a 12 x 12 x 12 Monkhorst-Pack k-point grid~\cite{Monkhorst76}. For the electron-ion interaction, we use a LDA-1/2 BHS norm-conserving scalar relativistic pseudopotential~\cite{Bachelet82, ferreira11}, for which the equilibrium lattice constant is $a_0=5.62$ \AA.} Table \ref{table-cG} gives the complex coefficients for $\bf G$ vectors such that $|c_{\mu} ({\bf G})|^2>10^{-3}$. All coefficients are given in the Supplementary Material~\cite{sup_mat}.
\begin{table}[ht!]
\centering
\begin{tabular}{ | c | c | c | c | c | c | }
 \hline
	$g_x$ & $g_y$ & $g_z$ &  Re$[c_1(\mathbf{G})]$ & Im$[c_1(\mathbf{G})]$ & $|c_1(\mathbf{G})|^2$ \\ \hline \hline
	1 & -1 & -1 & 0.320983& -0.132956 & 0.121 \\ \hline
	-1 & 1 & -1 &  0.320983& -0.132956 & 0.121 \\ \hline
	-1 & -1 & 1 &  0.320983& -0.132956 & 0.121 \\ \hline
	0 & 0 & -2 & 0.320983& 0.132956 & 0.121 \\ \hline
	0 & -2 & 0 & 0.320983& 0.132956 & 0.121 \\ \hline
	-2 & 0 & 0 & 0.320983 & 0.132956 & 0.121 \\ \hline
	0 & 0 & 0 & -0.110811 & 0.267521 & 8.38E-2 \\ \hline
	-1 & -1 & -1 & -0.110811 & -0.267521 & 8.38E-2 \\ \hline
	1 & 1 & -1 & 4.4389E-2 & 0.107165 & 1.35E-2 \\ \hline
	1 & -1 & 1 & 4.4389E-2 & 0.107165 & 1.35E-2 \\ \hline
	-1 & 1 & 1 & 4.4389E-2 & 0.107165 & 1.35E-2 \\ \hline
	0 & -2 & -2 & 4.4389E-2 & -0.107165 & 1.35E-2 \\ \hline
	-2 & 0 & -2 & 4.4389E-2 & -0.107165 & 1.35E-2 \\ \hline
	-2 & -2 & 0 & 4.4389E-2 & -0.107165 & 1.35E-2 \\ \hline
	1 & 1 & 1 & -6.6843E-2 & 2.7687E-2 & 5.23E-3 \\ \hline
	-2 & -2 & -2 & -6.6843E-2 & -2.7687E-2 & 5.23E-3 \\ \hline
\end{tabular}
\caption{Numerical values of the coefficients for the plane wave expansion of the periodic part of the Bloch function at one of the conduction band minima of Ge: $\mathbf{L_1}=\pi/a_0 (1, 1, 1)$, where $a_0=0.562$ nm is the equilibrium lattice constant.  Reciprocal lattice vectors are written in cartesian coordinates as $\mathbf{G} =  (2\pi/a_0) (g_x, g_y,g_z)$  where $g_{x,y,z}$ are given in the first 3 columns and the 4th column presents $|g| = \sqrt{g_x^2 + g_y^2 + g_z^2}$. Only plane waves with $|c_1(\mathbf{G})|^2>10^{-3}$ are shown. Coefficients for all other minima may be obtained from the symmetry relations:
$c_2 (i,j,k) = c_1^*(j,k,-i)$, $c_3 (i,j,k) = c_1^*(i,j,-k)$ and $c_4 (i,j,k) = c_1 (j,-k,-i)$, where $\{c_{\mu}\}$ are the coefficients for the minima $\mu =1,2,3,4$ at positions $\mathbf{L}_{\mu= 1,2,3,4}$, with $\mathbf{L_1}$ given above, $\mathbf{L_2}=(\pi/a_0)(-1, 1, 1)$, $\mathbf{L_3}=(\pi/a_0)(1,1,-1)$ and $\mathbf{L_4}=(\pi/a_0)(-1,1,-1)$. The minimum at each $L_{\mu}$ is equivalent by a reciprocal lattice translation to $-L_{\mu}$, and the coefficients at $-L_{\mu}$ are obtained from Kramers theorem as $c_{-L_\mu}(\mathbf{G})=c^*_{L_\mu}(-\mathbf{G})$.}
\label{table-cG}
\end{table}

\begin{figure}
\includegraphics[clip, width=0.49 \textwidth, natwidth=152mm, natheight=132mm]{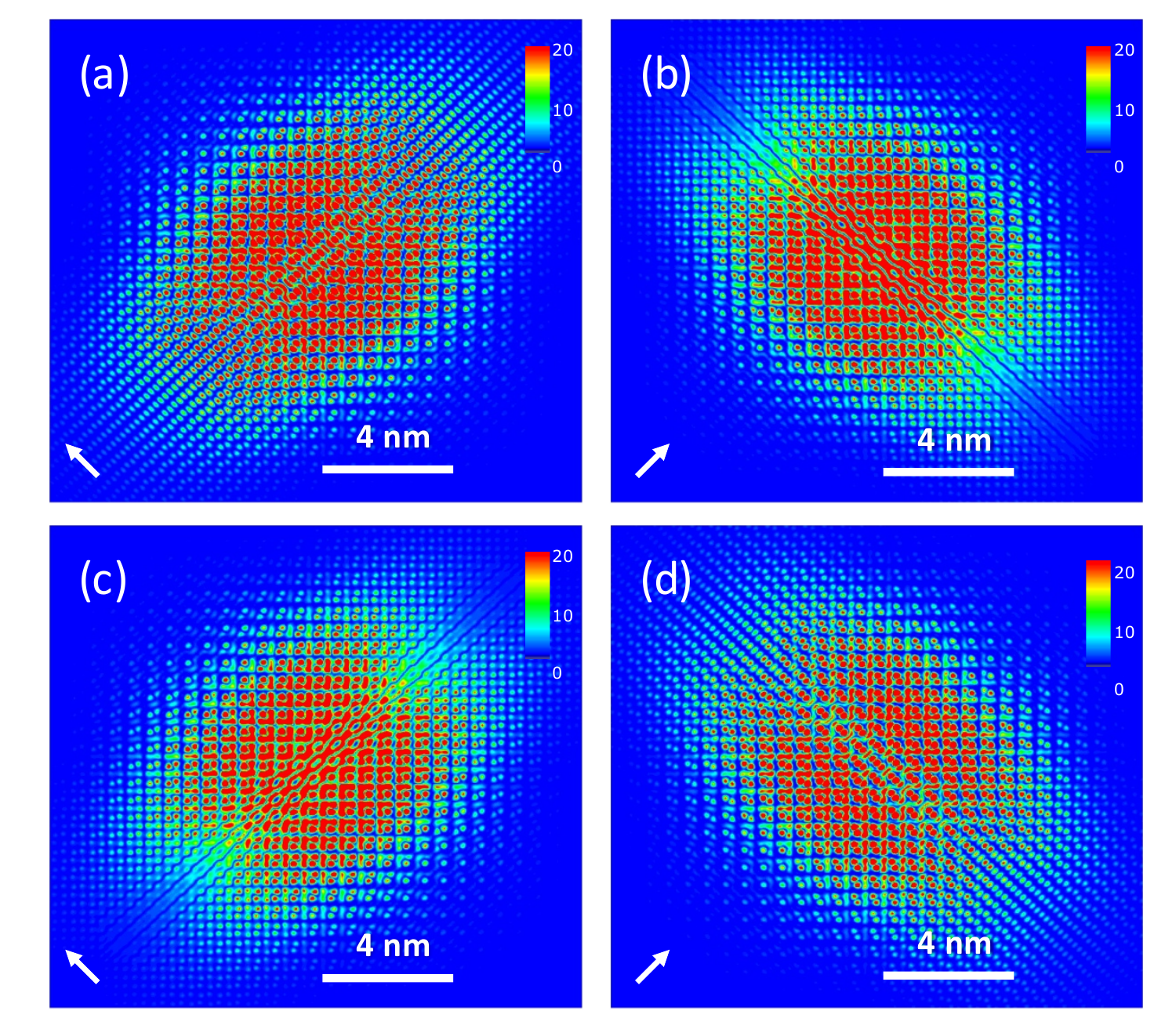}
\caption{STM images as predicted from interstitial planes cuts calculated from complete KL wavefunction for donors in Ge: Frames (a) to (d) correspond to $z=4 a_0 +j/4+a_0/8$ for $j=0, ..., 3$, respectively. The white arrows show the direction of the surface dimers which are perpendicular to the corresponding symmetry axis for each image. The color scale shows the magnitude of the charge density in units of $10^{-4}$ $(\rm a_{0}^{-3})$.}
\label{fig:STM-IP}
\end{figure}

Experimentally, donor images are collected by an STM tip kept at a fixed distance above the surface atomic plane. We fix the distance of the STM collected image to the surface to be $a_0/8$, so that the predicted image is a cut through the electronic density at intersticial planes mid-way between consecutive Ge atomic planes, located at heights $z=(n+j/4)a_0+a_0/8$ above the donor plane ($n$ Ge lattice parameters and $j$ monoatomic planes, defining 4 inequivalent relative positions of the interface plane with respect to the donor)~\cite{STM-si}. Interstitial plane cuts for the KL charge distributions for neutral P donors in Ge, in the sequence $j=0, 1, 2, 3$ for $n=4$, are given in Fig. \ref{fig:STM-IP}. These images should emulate images to be expected in STM signatures of sub-surface donors in Ge. Atomic plane cuts (not shown here) produce very different images, which are not observable for the STM tip standing above such planes.
There is a marked change in the patterns from atomic planes to interstitial planes.

The simulated images of burried donors in Ge (Fig. \ref{fig:STM-IP}) show elongated but almost featureless shapes, where the elongated axis is consistently perpendicular to the underlying dimer rows in Ge, alternating between the 2 possible dimerization directions. In comparison to donors in Si, where distinct shapes characterize successive monolayers, no noticeable differences appear between the four inequivalent relative positions of the Ge (001) surface plane and the impurity, namely $j=0,1,2,3$. Moreover, the images shown here for a given depth of the donor in Ge spread over a considerably larger area at the surface, with linear dimensions typically twice as those of Si.

All interstitial distributions are elongated in the  $[110]$ (frames a and c) or $[1\bar{1}0]$ (frames b and d) surface directions. This behavior is independent of the unit cell $n$: a $1\bar{1}1\bar{1}$ cyclic sequence results for interstitial planes along the $z$ direction, and is related to the cyclic directional orientation of the zig-zag atomic rows in the diamond structure~\cite{STM-si}. The larger spread of the donor electron in Ge in comparison with Si is a consequence of the lighter effective mass for an electron in Ge, thus larger Bohr radii (for example $a_{cc} = 2.89$ nm for P in Ge and $1.1$ nm for P in Si~\cite{Koiller2014}). Thus, comparatively, images in Ge appear fuzzy and featureless. A new feature that appears here is that the overall shape of the distributions remains constant for different depths of the substitutional donor with respect to the [100] surface. This is a consequence of the four L points being symmetrically oriented with the surface plane. In the case of Si, the $\Delta$ lines perpendicular to the surface are not equivalent to the parallel ones, resulting in shapes readily identifiable (butterfly and caterpillar in Ref. \onlinecite{STM-si}). One could expect differentiated images in Ge for [110] or [111] surfaces, which are not investigated here.

\section{Implications to Qubit Fabrication}\label{sec:discussions}

While Ge-based quantum computer design may benefit from the knowledge of multivalley electronic structure acquired by the Silicon qubit community, many quantitative aspects of the spectrum and wavefunction of dopants are unique to germanium. Its exceptionally large dielectric constant and low electronic effective mass lead to very spread charge distributions, which could potentially facilitate the tunneling of electron to and from such dopants. If this is the main task for a given qubit design, our calculations indicate that Sb donors are particularly suitable, with a wavefunction spreading over more than 4 nm.

On the other hand, the binding of a second electron to the dopant, forming a D$^-$ state, is not naturally strong (hundreds of $\mu$eV). This indicates that this state should not be stable in bulk at high temperatures, but can be made stable in nanostructures (specially near metals where the image charge generates some attractive potential~\cite{calderon}) or by implanting donors by pairs~\cite{gonzalez-zalba}. This is important in some proposals for spin manipulation based on spin-charge conversion. Donors in two-electron configurations have also been proposed to promote entanglement between nucleus-electron spin qubits~\cite{kane-prl-2003}. These ideas should be easier to implement with arsenic donors for a stronger binding of the second electron.

The wavefunction oscillation pattern due to the interference among Bloch states of the conduction band minima in Ge leads to a unique signature that might be exploited to identify these dopants from STM measurement. Yet, the fast oscillations might impair the use of these images for a complete three-dimensional determination of the donor position -- unlike in silicon, where the charge distribution presents coarse signatures of the sublattice in which the substitutional donor is placed. A finer comparison between the oscillatory pattern from theory and the tunneling current might overcome this limitation, but an even more accurate description of the wavefunction may be required.

These physical attributes indicate that germanium can be superior to silicon for some tasks required for quantum applications, but not all. It is clear, though, that many procedures conceived for silicon architectures are readily transferable to Ge devices, which should stimulate further experimental progress for Ge-based qubits.

\section{Acknowledgments} This work was performed as part of the Brazilian National Institute for Science and Technology on Quantum Information. We also acknowledge partial support from the Brazilian agencies FAPERJ, CNPq, CAPES.

%\newpage

\bibliography{Ge-biblio}

\end{document}